%% file: main.tex
\documentclass[12pt]{article}
\usepackage[utf8]{inputenc}
\usepackage{amsmath, amsthm, amssymb, amsfonts}
\usepackage{import}
\usepackage[numbers, comma]{natbib}
\usepackage{fullpage}
\usepackage{authblk}
\usepackage{apxproof}
\usepackage{import}
\usepackage{mathtools}
\usepackage{tikz}
\usepackage{float}
\usepackage{enumitem}

\usetikzlibrary{calc,patterns,angles,shapes, positioning, intersections, quotes}

\newtheorem{theorem}{Theorem}
\newtheoremrep{thm}{Theorem}
\newtheoremrep{lem}{Lemma}

\usepackage{graphicx}
\newtheorem{corollary}[theorem]{Corollary}

\newtheorem{conjecture}[theorem]{Conjecture}

\newtheorem{remark}[theorem]{Remark}
\theoremstyle{definition}
\newtheorem{definition}{Definition}[section]

\usepackage[colorlinks=false,pagebackref=true, hidelinks]{hyperref}

\newcommand{\R}{\mathbb{R}}
\def\G{\Gamma}
\def\a{\alpha}

\def\x{\vec{x}}
\let\vec\mathbf
\DeclarePairedDelimiter{\norm}{\lVert}{\rVert} 

\date{}
\allowdisplaybreaks

\begin{document}
\title{Optimality of the coordinate-wise median mechanism for strategyproof facility location in two dimensions \thanks{We are grateful to Arunava Sen, Federico Echenique, Tom Palfrey, Omer Tamuz, Debasis Mishra for helpful comments and suggestions. An earlier version of this paper circulated under the title "Coordinate-wise median: Not bad, Not bad, Pretty good." }}
\author{Sumit Goel\thanks{California Institute of Technology; sgoel@caltech.edu; 0000-0003-3266-9035} \quad Wade Hann-Caruthers \thanks{California Institute of Technology; whanncar@gmail.com; 0000-0002-4273-6249}}

\maketitle

\import{files/}{abstract.tex}

\import{files/}{intro.tex}

\import{files/}{model.tex}

\import{files/}{best.tex}

\import{files/}{minisum.tex}

\import{files/}{p-norm.tex}

\import{files/}{conclusion.tex}

\newpage
\nocite{*}

\bibliographystyle{ecta}

\bibliography{refs}

\end{document}

%% file: files/abstract.tex
\begin{abstract}
We consider the facility location problem in two dimensions. In particular, we consider a setting where agents have Euclidean preferences, defined by their ideal points, for a facility to be located in $\mathbb{R}^2$.  We show that for the  $p-norm$ ($p \geq 1$) objective, the coordinate-wise median mechanism (CM) has the lowest worst-case  approximation ratio  in the class of deterministic, anonymous, and strategyproof mechanisms. For the minisum objective and an odd number of agents $n$, we show that CM has a worst-case approximation ratio (AR) of $\sqrt{2}\frac{\sqrt{n^2+1}}{n+1}$. For the $p-norm$ social cost objective ($p\geq 2$),  we find that the AR for CM is bounded above by $2^{\frac{3}{2}-\frac{2}{p}}$. We conjecture that the AR of CM actually equals the lower bound $2^{1-\frac{1}{p}}$ (as is the case for $p=2$ and $p=\infty$) for any $p\geq 2$. 

\end{abstract}

%% file: files/intro.tex
\section{Introduction}
We consider the problem of locating a facility on a plane where a set of strategic agents have private preferences over the facility location. Each agent's preference is defined by its ideal point so that the cost incurred by an agent equals the Euclidean distance between the facility location and the ideal point. A central planner wishes to locate the facility to minimize the social cost. Since agents may lie about their ideal points if it benefits them, the planner is constrained to choose a mechanism that is strategyproof. In this paper, we consider the problem of finding the strategyproof mechanism that best approximates the optimal social cost as measured by the worst-case approximation ratio (AR) and quantifying its performance.\\

We find that for an odd number of agents $n$ and the $p-norm$ objective with $p\geq 1$, the coordinate-wise median mechanism is optimal in the class of deterministic, anonymous, and strategyproof mechanisms. For the utilitarian social objective of minimizing the sum of individual costs ($p=1$), we show that the coordinate-wise median mechanism has an AR of $\sqrt{2}\frac{\sqrt{n^2+1}}{n+1}$. For the general $p-norm$ objective $(p\geq 2)$, we show that the asymptotic AR of the coordinate-wise median mechanism is bounded between $2^{1-\frac{1}{p}}$ and $2^{\frac{3}{2}-\frac{2}{p}}$. We conjecture that the asymptotic AR of the coordinate-wise median mechanism is actually equal to the lower bound $2^{1-\frac{1}{p}}$ (as is the case when $p=2$ or $p=\infty$). \\

This problem has been extensively studied in the literature known as \textit{Approximate Mechanism Design without money}. It was first introduced by  \citet{procaccia_approximate_2013} who studied the setting of locating a single facility on a real line under the utilitarian (sum of individual costs) and egalitarian (maximum of individual costs) objectives. Since then, the problem has received much attention, with extensions to alternative objective functions, multiple facilities, obnoxious facilities, different networks, etc. \citet{gao_survey_2015} and more recently,  \citet{chan_mechanism_2021} provide surveys of results in the last decade in several of these settings. In the class of deterministic strategyproof mechanisms for locating a facility, the median mechanism has been shown to be optimal under various objectives and domains [\citet{procaccia_approximate_2013}, \citet{feigenbaum_approximately_2017}, \citet{feldman_strategyproof_2013}, \citet{feldman_voting_2016}].  \\

There has been some related work in extending the problem to multiple dimensions. \citet{meir_strategyproof_2019} shows that in the d-dimensional Euclidean space, the approximation ratio of the coordinate-wise median mechanism for the utilitarian objective is bounded above by $\sqrt{d}$.  \citet{sui_analysis_2013} propose percentile mechanisms for locating multiple facilities in Euclidean space which are further analysed in  \citet{sui_approximately_2015} and  \citet{walsh_strategy_2020}.
 \citet{meir_strategyproof_2019}, using techniques different from ours, finds the AR of coordinate-wise median mechanism under the minisum objective for the case of $3$ agents.  \citet{gershkov_voting_2019} shows that for some natural priors on the ideal points (that include i.i.d. marginals), taking the coordinate-wise median after a judicious rotation of the orthogonal axes can lead to welfare improvements under the least-squares objective. In other related work, \citet{el-mhamdi_strategyproofness_2021} find that the mechanism choosing the minisum optimal location (geometric median) is approximately strategyproof in a large economy. \citet{lee_brady_spatial_2016} find that the geometric median is Nash-implementable and in the case of three agents, it  is the unique rule that satisfies anonymity, neutrality, and Maskin-Monotonicity. \citet{durocher_projection_2009} and \citet{bespamyatnikh_mobile_2000} analyse   approximations to geometric median due to its instability and computational difficulty.
% who  considers facility location with mobile agents,   group strategyproofness [\citet{sui_analysis_2013}, \citet{sui_approximately_2015}, \citet{walsh_strategy_2020}, \citet{durocher_projection_2009}, \citet{bespamyatnikh_mobile_2000}]. 
\\

There is also a large literature in social choice theory on characterizing the set of strategyproof mechanisms under different assumptions on preference domains [\citet{gibbard_manipulation_1973}, \citet{satterthwaite_strategy-proofness_1975}, \citet{moulin_strategy-proofness_1980}]. In multiple dimensions with Euclidean preferences, the characterizations typically include or are completely described by the coordinate-wise median mechanism [\citet{kim_nonmanipulability_1984}, \citet{border_straightforward_1983}, \citet{peters_range_1993}, \citet{peters_pareto_1992}]. Our work augments this literature, which provides strong axiomatic foundations for the coordinate-wise median mechanism, by demonstrating its quantitative optimality.\\

The paper proceeds as follows. In section 2, we formally define the problem and state some characterisation results and approximation results from the literature that will be useful in our analysis. In section 3,  we discuss the optimality of the coordinate-wise median mechanism. In sections 4 and 5, we discuss the problem of finding the approximation ratio of the coordinate-wise median mechanism for the utilitarian objective and the $p$-norm objective. Section 6 concludes.\\

%% file: files/model.tex
\section{Preliminaries}

Suppose $(X,d)$ is a metric space. There are $n$ agents and each agent has an ideal point  $x_i \in X$ for a facility to be located in $X$. The cost of locating the facility at $y \in X$ for agent $i$ is $d(y,x_i)$. Let $\x$ be the profile of ideal points: $\x = (x_1, \dots, x_n)$.  The social cost of locating the facility at $y$ under profile $\x \in X^n$ is given by the \textit{social cost function} $sc : X \times X^n \to \R$. Let $OPT(sc, \vec{x})$ denote the set of minimizers for $sc$ given $\vec{x}$:
\[OPT(sc, \vec{x}) = \text{argmin}_y sc(y, \vec{x}).\]
When $OPT(sc, \cdot)$ is singleton-valued, we will abuse notation and use $OPT(sc, \x)$ to refer to the unique element contained therein. When $sc$ is clear from context, we will suppress the first argument and write $OPT(sc, \vec{x})$ simply as $OPT(\vec{x})$. \\

A \textit{mechanism} is a function $f : X^n \to X$. It is said to be \textit{strategyproof} if no agent can benefit by misreporting her ideal point, regardless of the reports of the other agents. Formally:
\begin{definition}
A mechanism $f$ is strategyproof if for all $i \in N$, $x_i, x_i' \in X$, $x_{-i} \in X^{n-1}$, \[d(f(x_i,x_{-i}), x_i) \leq d(f(x_i',x_{-i}), x_i).\]
\end{definition}

\begin{definition}
A mechanism $f$ is anonymous if for any permutation $\pi: [n] \to [n]$, 
$$f(x_1, \dots, x_n)=f(x_{\pi(1)}, \dots, x_{\pi(n)})$$
\end{definition}

To measure how closely a mechanism approximates the optimal social cost for a given profile, we use the approximation ratio.

\begin{definition}
For a social cost function $sc$, the approximation ratio of a mechanism $f$ at a profile $\x$ is given by
\begin{align*}
    AR_f(\x) = \frac{sc(f(\x), \x)}{sc(OPT(\x), \x)}.
\end{align*}
\end{definition}
In the case that $sc(OPT(\x), \x) = 0$, we take $AR_f(\x)$ to be $1$ if $sc(f(\x), \x) = 0$ and $\infty$ otherwise.

To compare mechanisms, we will evaluate them by their worst-case approximation ratio.

\begin{definition}
The worst-case approximation ratio of a mechanism $f$ is given by
\begin{align*}
    AR(f) = \sup_{\x}{AR_f(\x)}.
\end{align*}
\end{definition}

Given a metric space $(X,d)$ and a social cost function $sc$, the problem is to find a strategyproof mechanism with the smallest worst-case approximation ratio.\\

In this paper, we consider the Euclidean metric space with $X=\R^2$ and the $p$-norm social cost function which is the $L_p$ norm of the vector of Euclidean distances $sc(y,\vec{x})=\left[\sum \norm{y-x_i}^p\right]^\frac{1}{p}$ where $p\geq 1$. We refer to the coordinates of points in $\R^2$ by $a$ and $b$. We refer to the sets $\R \times \{0\}$ and $\{0\} \times \R$ as the $a$-axis and the $b$-axis, respectively. We refer to the sets $\pm \R_{\geq 0} \times \{0\}$ and $\{0\} \times \pm \R_{\geq 0}$ as the $\pm a$-axes and $\pm b$-axes, respectively. We use the notation $[y,z]$ to denote the line segment joining $y$ and $z$: $\{t y + (1 - t) z \, : \, t \in [0, 1]\}$. Similarly, we denote by $(y, z)$ the set $[y,z] \setminus \{y,z\}$.\\

Our analysis makes use of some previous results regarding characterization of strategyproof mechanisms and bounds on approximation ratios  in the Euclidean domain.  We collect those results here.\\

\subsection{Characterisation results in two dimensions}

First, let's define an important class of mechanisms in this domain. 

\begin{definition}
In the Euclidean metric space with $X=\R^m$, a mechanism $f$ is called a \textit{generalized coordinate-wise median mechanism} with $k$ constant points if there exists a coordinate system and points $c_{1}, c_{2}, \ldots, c_{k} \in(\R \cup\{-\infty, \infty\})^{m}$ so that for every profile $\x \in\left(\R^{m}\right)^{n}$ and every dimension $j=1,2, \ldots, m$, the $j^{th}$ coordinate of $f$ is given by
$$
f^{j}(\x):=\operatorname{med}\left(x_1^{j}, x_2^{j}, \ldots, x_n^{j}, c^{j}_{1}, \ldots, c^{j}_{k}\right)
$$
where``med" denotes the median of the subsequent real numbers.
\end{definition}

This class of mechanisms has strong axiomatic foundations in the literature as illustrated in the following lemma:

\begin{lem}[\citet{kim_nonmanipulability_1984,peters_pareto_1992,peters_range_1993}]
\label{characterisation}
In the Euclidean metric space with $X=\R^2$ and an odd number of agents $n$, a mechanism $f:\left(\mathbb{R}^{2}\right)^{n} \rightarrow \mathbb{R}^2$ is
\begin{itemize}
    \item (\citet{kim_nonmanipulability_1984})  continuous, anonymous, and strategyproof if, and only if,  $f$ is a generalized coordinate-wise median mechanism with $n+1$ constant points.
    \item (\citet{peters_range_1993}) unanimous, anonymous, and strategyproof if, and only if,  $f$ is a generalized coordinate-wise median mechanism with $n-1$ constant points.
    \item (\citet{peters_pareto_1992}) Pareto optimal, anonymous, and strategyproof if, and only if, $f$ is a generalized coordinate-wise median mechanism with $0$ constant points.
\end{itemize}
\end{lem}

We refer to the generalized coordinate-wise median mechanism with $0$ constant points and the standard coordinate-system as the \textit{coordinate-wise median mechanism} and  denote it by $c(\vec{x})$.

One subclass of generalized coordinate-wise median mechanisms that will play an important role in demonstrating the optimality of the coordinate-wise median mechanism is the following:

\begin{definition}
In the Euclidean metric space with $X=\R^m$, a mechanism $f$ is called a \textit{coordinate-wise quantile mechanism} if it is a generalized coordinate-wise mechanism where all the $k$ constant points $c_{1}, c_{2}, \ldots, c_{k} \in \{-\infty, \infty\}^{m}$.
\end{definition}

Note that if $|\{ i \, : \, c_i^j = -\infty\}| = \ell$, then $f^j(x_1, \dots, x_n)$ is the $\frac{n+k+1}{2} - \ell$ order statistic of the (multi)set $\{x_1^j, \cdots, x_n^j\}$. Hence, given a profile $\x$, every coordinate-wise median quantile mechanism locates the
facility by selecting, for each dimension $j$, some fixed quantile of the ordered projection of $\x$ in the $j^{th}$ dimension as
the coordinate of the facility location.

\subsection{Approximation results in two dimensions}
For the case of $X=\R^2$ with the Euclidean metric, there has been some work in finding bounds on AR for the utilitarian objective $sc(y, \vec{x})=\sum \norm{y-x_i}$. We discuss those findings here.

A point minimizing the sum of distances from a finite set of points in $\R^2$ is known as a \textit{geometric median} for that set of points. The geometric median is characterised by the following result:

\begin{lem}
\label{lem:gm}
Given $\vec{x} \in (\R^2)^n$, a point $y \in \R^2$ is a geometric median for $\vec{x}$ if and only if there are vectors $u_1, \dots, u_n$ such that $$\sum_{i=1}^{n} u_i=0$$ where for $x_i \neq y$, $u_i=\frac{x_i-y}{\norm{x_i-y}}$ and for $x_i=y$, $\norm{u_i} \leq 1$.
\end{lem}

This characterisation yields conditions under which changing a profile of points does not change the geometric median, as summarized in the following corollary:

\begin{corollary}
\label{cor:gm}
Let $\vec{x}\in (\R^2)^n$, and denote by $y$ the geometric median of $\vec{x}$. For any $i$, if $x_i \neq y$ and if $x_i' \in  \{ y + t(x_i - y) \, | \, t \in \R_{\geq 0} \} $, then the geometric median for the profile $(x_i', x_{-i})$ is also $y$.
\end{corollary}
Informally, moving a point directly away from or directly towards (but not past) the geometric median leaves the geometric median unchanged. We will use this observation repeatedly in the sequel and note here that in fact it will be the only characteristic of the geometric median that we use for much of the paper. We refer to the geometric median by $g(\vec{x})$.\\

% \footnote{The geometric median is unique whenever $n$ is odd or points are not collinear.}.

It follows from Lemma~\ref{characterisation} that the geometric median mechanism is not strategyproof.  \citet{meir_strategyproof_2019} finds an upper bound on the AR of the coordinate-wise median mechanism in the $m-$dimensional problem:
\begin{lem}[\citet{meir_strategyproof_2019}]
\label{bound}
For $X=\R^m$ and the utilitarian  objective $sc(y,\vec{x})=\sum \norm{y-x_i}$, the coordinate-wise median mechanism has an approximation ratio of at most $\sqrt{m}$ for any number of agents $n$.
\end{lem}

% The proof follows from observing that the median is optimal in one dimension and that for any right triangle, the sum of the lengths of the legs is at most $\sqrt{m}$ times the length of the hypotenuse. 

% \citet{bespamyatnikh_mobile_2000} find that the WAR for $c(\x)$ is bounded above by $\sqrt{2}$. 

% \citet{durocher_projection_2009}, motivated by the instability of geometric median, consider the problem of finding other mechanisms that best approximate the minimum sum while being stable. They show that the upper bound on WAR of $c(\vec{x})$ is tight.\\

% \footnote{It is unstable in the sense that small perturbations in inputs can result in arbitrarily large change in the position of the median. For example: $\x=\{(0,0), (0,0), (1,0), (1,\epsilon)\}$ and $\x'=\{(0,0), (0, \epsilon), (1,0), (1,0)\}$}

\citet{feigenbaum_approximately_2017} consider the facility location problem for $X=\R$ and $d(x_i,x_j)=|x_i-x_j|$ with the social cost function $sc(y,\x)=\left[\sum |y-x_i|^p\right]^\frac{1}{p}$. 

\begin{lem}[\citet{feigenbaum_approximately_2017}]
\label{1d}
For $X=\R$ and the $p$-norm objective $sc(y,\x)=\left[\sum |y-x_i|^p\right]^\frac{1}{p}$ with $p \geq 1$, the median mechanism has an approximation ratio of  $2^{1-1 / p}$. Further, any deterministic strategyproof mechanism has approximation ratio of at least $2^{1-1 / p}$.
\end{lem}

% \subsection{Preview of results}

% For the case of $p=1$, we show that the approximation ratio for $c(\x)$ is  $\frac{\sqrt{2}\sqrt{n^2+1}}{n+1}$ when there are $n$ agents. This bound increases to $\sqrt{2}$ as $n \to \infty$. We further show that no other deterministic, anonymous and strategyproof mechanism can have a better approximation ratio for the utilitarian objective.
% For $p\geq 2$, we show that approximation ratio of $c(\x)$ is bounded above by $2^{\frac{3}{2}-\frac{2}{p}}$. Together with the corollary, the bound implies that $c(\x)$ is very close to being optimal. For $p=2$ and $p=\infty$, the approximation ratio of coordinate-wise median is actually equal to the lower bound in corollary \ref{lb} which leads us to conjecture that it is true more generally for any $p \geq 2$. If the conjecture is true and AR of $c(\vec{x})$ is indeed  $2^{1-\frac{1}{p}}$, it would follow from Corollary ~\ref{lb} that it is the best deterministic SP mechanism.

%% file: files/best.tex
\section{Optimality of the coordinate-wise median mechanism}

% Sketch: 

%  + Theorem statement: CM optimal with p-norms and minimax
%  + Show that for any affine social welfare function, for any mechanism there must be a better quantile mechanism
%  + Show that for any convex social welfare function, any quantile mechanism has an approximation ratio at least as big as the coordinatewise median's
 
Our first major finding is that the coordinate-wise median mechanism is optimal with respect to the worst-case approximation ratio for the class of social cost functions we study.

\begin{thm}
\label{thm:cm-best}
For $X=\R^2$, and the $p$-norm objective $sc(y, \vec{x})=\left[\sum \norm{y-x_i}^p \right]^\frac{1}{p}$ where $p\geq 1$, the coordinate-wise median mechanism has the lowest approximation ratio among all deterministic, anonymous, and strategyproof mechanisms.
\end{thm}

To prove Theorem~\ref{thm:cm-best}, we will show that for every deterministic, anonymous, and strategyproof mechanism $f$, there is a coordinate-wise quantile mechanism $Q$ such that $AR(f) \geq AR(Q) \geq AR(CM)$. In the case that $f$ is not unanimous, $AR(f) = \infty > AR(CM)$. In the case that $f$ is unanimous, it follows from Lemma~\ref{characterisation} that $f$ is a generalized coordinate-wise median mechanism with $n-1$ constant points. Thus, to prove the theorem, we will show that for every such mechanism there is a coordinate-wise quantile mechanism with a lower AR (Lemma~\ref{lem:quantiles-optimal}) and that CM has the lowest AR among all coordinate-wise quantile mechanisms (Lemma~\ref{lem:cm-optimal-among-quantiles}).

\begin{lem}
\label{lem:quantiles-optimal}
Let $f$ be a generalized coordinate-wise median mechanism with $n-1$ constant points. Then for any $p$-norm objective $sc$, there is some coordinate-wise quantile mechanism $Q$ such that $AR(f) \geq AR(Q)$.
\end{lem}
\begin{proof}
Let $c_1, \dots, c_{n-1}$ be the constant points for $f$. Let $Q$ be the coordinate-wise quantile mechanism with constant points $q_1, \dots, q_{n-1}$, where for each $i$, $q_i^j = \infty$ if $c_i^j = \infty$ and $q_i^j = -\infty$ otherwise. Now we'll show that $AR(f) \geq AR(Q)$. 

Let $z^1 \in \R$ such that for every $i$, either $c_i^1 < z^1$ or $c_i^1 = \infty$, and similarly, let $z^2 \in \R$ such that for every $i$, either $c_i^2 < z^2$ or $c_i^2 = \infty$. Then for any $\x$ such that $x_i^1 > z^1$ and $x_i^2 > z^2$ for all $i$, it follows immediately from the definition of the $q_i$ that
\begin{align*}
    med(x_1^j, \dots, x_n^j, q_1^j, \dots, q_{n-1}^j) = med(x_1^j, \dots, x_n^j, c_1^j, \dots, c_{n-1}^j).
\end{align*}
Defining $T = ([z^1, \infty] \times [z^2, \infty])^n$, it hence follows that $f(\x) = Q(\x)$ for all $\x \in T$, and thus that $AR_Q(\x) = AR_f(\x)$ for all $\x \in T$. 

In addition, we note that for any $\x \in (\R^2)^n$, if $\x'$ is a translation of $\x$ (i.e. there is some $\Delta x \in \R^2$ such that $x_i' = x_i + \Delta x$ for all $i$), then $AR_Q(\x) = AR_Q(\x')$.

Putting these observations together, it then follows that
\begin{align*}
    AR(Q) &= \sup_{\x \in (\mathbb{R}^2)^N} AR_Q(\x)\\ 
    &= \sup_{\x \in T} AR_Q(\x)\\ 
    &= \sup_{\x \in T} AR_f(\x)\\
    &\leq \sup_{\x \in (\mathbb{R}^2)^N} AR_f(\x)\\
    &= AR(f).
\end{align*}
\end{proof}

\begin{lem}
\label{lem:cm-optimal-among-quantiles}
Let $Q$ be a coordinate-wise quantile mechanism. Then for any $p$-norm objective $sc$, $AR(Q) \geq AR(CM)$.
\end{lem}

\begin{proof}
Since $Q=Q_1$ is a coordinate-wise quantile mechanism, there exist order statistics $(k_1,k_2) \in [n] \times [n]$ such that $Q$ locates the facility by selecting, for each dimension, the $k_i$th order statistic of the projection of $\x$ in the $i$th dimension as
the coordinate of the facility in the $i$th dimension.

Now consider the coordinate-wise quantile mechanisms $Q_2,Q_3,Q_4$ defined by the order statistics $(k_1,n+1-k_2), (n+1-k_1,k_2), (n+1-k_1,n+1-k_2)$ respectively. As these mechanisms are isomorphic (each can be obtained from the others by composition with a series of reflections across axes), they all have the same approximation ratio; that is, $AR(Q_i) = AR(Q)$ for each $i$. Observe that for any profile $\x \in (\mathbb{R}^2)^n$, $CM(\x)$ is in the convex hull of $Q_1(\x),Q_2(\x)$ $,Q_3(\x),Q_4(\x)$. Hence, since $sc(\x, y)$ is quasi-convex as a function of $y$,\footnote{Since $sc(\x, y)^p$ is convex as a function of $y$ and the composition of a convex function with a nondecreasing function is quasi-convex, $sc(\x, y) = (sc(\x, y)^p)^\frac{1}{p}$ is quasi-convex as a function of $y$.} $sc(\x, CM(x)) \leq sc(\x, Q_\ell(x))$ for some $\ell \in {1, 2, 3, 4}$, and so $AR_{CM}(\x) \leq AR_{Q_\ell}(\x) \leq AR(Q_\ell) = AR(Q)$. Thus, $AR_{CM}(\x) \leq AR(Q)$ for all $\x$, and so $AR(CM) \leq AR(Q)$.
\end{proof}

\begin{remark}
The techniques used to prove Theorem~\ref{thm:cm-best}, together with characterization results for the one-dimensional facility location problem (\citet{moulin_strategy-proofness_1980}), can also be used to prove that the median mechanism is also optimal for any $n$ odd and any $p \geq 1$. This strengthens the result in \citet{feigenbaum_approximately_2017} (Lemma \ref{1d}), which demonstrates there is no mechanism that is \textit{asymptotically} superior to the median mechanism.
\end{remark}

\begin{remark}
Lemma~\ref{lem:quantiles-optimal} and Lemma~\ref{lem:cm-optimal-among-quantiles} both hold for larger classes of social cost functions than $p$-norms. For Lemma~\ref{lem:quantiles-optimal}, it is sufficient that $sc$ depends only on the distances to the facility, and for Lemma~\ref{lem:cm-optimal-among-quantiles}, it is sufficient that $sc$ is quasiconvex. It follows that Theorem~\ref{thm:cm-best} holds for a much more general class of social cost functions. For instance, it holds when the planner's objective is to minimize a weighted sum of distances $sc(y,\x)=\sum_{i=1}^n \lambda_i \norm{y-x_i}$.
\end{remark}

%% file: files/minisum.tex
\section{The minisum objective}

% \wade{We need to fix the a's and b's.}

In this section, we quantify exactly the approximation ratio for the coordinate-wise median mechanism under the minisum ($p=1$) objective $sc(y,\vec{x})=\sum_{i=1}^n \norm{y-x_i}$. By Theorem~\ref{thm:cm-best}, it follows that this quantity provides a lower bound for the approximation ratio of any deterministic, anonymous, and strategyproof mechanism under the minisum objective.

\begin{thmrep}
\label{minisum}
For $n$ odd, $X=\mathbb{R}^2$, and  $sc(y,\vec{x})=\sum_{i=1}^n \norm{y-x_i}$, 
% the worst-case approximation ratio for the coordinate-wise median mechanism is given by: 
$$ AR(CM)=\sqrt{2}\dfrac{\sqrt{n^2+1}}{n+1}.$$
\end{thmrep}

\begin{appendixproof}

Define Centered Perpendicular (CP) profiles as all profiles $\vec{x} \in (\mathbb{R}^2)^n$ such that 

\begin{itemize}
    \item $c(\vec{x})=(0,0)$
    \item for all $i$, either $a_i=0$ or $b_i=0$ or $x_i=g(\vec{x})$
    \item if $x_i' \in (x_i, g(\vec{x}))$, then $c(x_i', x_{-i}) \neq (0,0)$
\end{itemize}

\begin{lem}[CP]
\label{cpn}
For any profile $\vec{x} \in (\mathbb{R}^2)^n$, there exists a profile $\vec{\chi} \in CP$ such that $AR(\vec{\chi}) \geq AR (\vec{x})$.
 \end{lem}
 
 \begin{proof}
 Let $\vec{x} \in (\mathbb{R}^2)^n$ be a profile. Let $\vec{x}'$ be the profile where $x_i' = x_i - c(\vec{x})$. Then $\vec{x}'$ has the same approximation ratio and $c(\vec{x}') = (0,0)$. Denote $A = \{i: a_i=0\}$ and $B = \{i: b_i=0\}$. Note that since $c(\vec{x}') = (0, 0)$, it follows from the definition of $c(\vec{x}')$ that $A \neq \emptyset$ and $B \neq \emptyset$. Let $\G = \{(a, b) \, : \, a = 0 \text{ or } b = 0\} \cup g(\vec{x}')$. Starting from  $i=1$ and going till $n$, define $x_i''$ to be the point in $[x_i', g(\vec{x}')] \cap \G$ that is closest to $g(\vec{x}')$ under the constraint that $c(x_1'', x_2'', \dots, x_i'', x_{i+1}, x_n)=(0,0)$.  Then  $\vec{x}'' \in CP$. Further, by lemma \ref{togm} $AR(\vec{x}'') \geq AR(\vec{x}') = AR(\vec{x})$; hence, taking $\vec{\chi} = \vec{\vec{x}''}$ completes the proof.
 \end{proof}

 Define Isosceles-Centered Perpendicular (I-CP) profiles as all $\vec{x} \in CP$ for which there exists $t \geq 0$ such that
\begin{itemize}
    \item $x_1= \dots =x_m=(t,0)$
    \item $x_{m+1}=(-t, 0)$
    \item $x_{m+2}= \dots= x_{2m+1}=(0,1)$
    \item $g(\vec{x})=(0,1)$.
\end{itemize}

Next, we prove some lemmas that will be useful in reducing the search space for the worst-case profile from $CP$ to $I-CP$.

First, we show that we can reduce the number of half-axes that the points lie on from (at most) four to (at most) three.

\begin{lem}[Reduce axes]
\label{Reduce axes}
Suppose $\vec{x}$ and $\vec{x}'$ are profiles which differ only at $i$ where for some $a > 0$, $x_i = (0, -a)$ and $x_i' = (-a, 0)$, and for which $c(\vec{x}) = c(\vec{x}') = (0, 0)$ and $b_g(\vec{x}) \geq a_g(\vec{x}) \geq 0$. Then $AR(\vec{x}') \geq AR(\vec{x})$.
\end{lem}

\begin{proof}
Again $c(\vec{x}')=c(\vec{x})$ and $sc(c(\vec{x}'), \vec{x}')=sc(c(\vec{x}), \vec{x})$. Thus, it is sufficient to show that $sc(g(\vec{x}'), \vec{x}') \leq sc(g(\vec{x}), \vec{x})$. For this, we just need to show that $d(x_i', g(\vec{x})) \leq d(x_i, g(\vec{x}))$. This follows from the following simple calculation:
\begin{align*}
    d(x_i', g(\vec{x}))^2 &= (a_g(\vec{x}) + a)^2 + b_g(\vec{x})^2\\
    &= a_g(\vec{x})^2 + 2 a_g(\vec{x})a + a^2 + b_g(\vec{x})^2\\
    &\leq a_g(\vec{x})^2 + b_g(\vec{x})^2 + 2a b_g(\vec{x}) + a^2\\
    &= a_g(\vec{x})^2 + (b_g(\vec{x}) + a)^2\\
    &= d(x_i, g(\vec{x}))^2.
\end{align*}
\end{proof}

Next, we show that we can combine points on each of the three half-axes while weakly increasing the approximation ratio.

\begin{lem}[Convexity]
\label{convexity}
Let $\vec{x} \in CP$ and let $S \subseteq N$ be such that for all $i \in S$, $a_i > 0$ and $b_i = 0$. Let $x_S$ be the mean of the $x_i$ across $i \in S$. Let $\vec{x}'$ be the profile where
\begin{enumerate}
    \item $x_j' = x_j$ for $j \notin S$ and
    \item $x_j' = x_S$ for $j \in S$.
\end{enumerate}
Then $AR(\vec{x}') \geq AR(\vec{x})$.
\end{lem}

\begin{proof}
It is immediate that $c(\vec{x}') = c(\vec{x})$. Hence, it will be sufficient to show that AR for $\vec{x}'$ with $c(\vec{x})$ and $g(\vec{x})$ instead of $c(\vec{x}')$ and $g(\vec{x}')$ is at least as big as $AR(\vec{x})$.
Indeed,  $sc(c(\vec{x}),\vec{x}')=sc(c(\vec{x}),\vec{x}))$ and $sc(g(\vec{x}'),\vec{x}') <sc(g(\vec{x}),\vec{x}') <sc(g(\vec{x}),\vec{x})$ where the last inequality follows from convexity of the distance function.
\end{proof}

The same argument applies for any of the other strict half axes.

Next, we show that we can move all the points that are on the geometric median to the axis in a way that weakly increases the approximation ratio.

\begin{lem}[Double Rotation]
\label{Double Rotation}
Let $\vec{x}$ and $\vec{x}'$ be profiles that differ only at $i_1$ and $i_2$, such that for some $a \geq 0$
\begin{itemize}
    \item $c(\vec{x}) = (0, 0)$,
    \item $b_g(\vec{x}) \geq a_g(\vec{x}) > 0$,
    \item $x_{i_1} = (-a, 0)$,
    \item $x_{i_1}' = (a + 2 a_g(\vec{x}), 0)$,
    \item $x_{i_2} = g(\vec{x})$, and 
    \item $x_{i_2}' = (0 , d(g(\vec{x}), (0, 0)))$.
\end{itemize} 
Then $c(\vec{x}') = (0, 0)$ and $AR(\vec{x}') \geq AR(\vec{x})$.
\end{lem}

\begin{proof}
The first claim is immediate.

For the second claim, let
\begin{align*}
    A &= \sum_{i \neq i_1}{d(x_i, c(\vec{x}))}\\
    B &= \sum_{i \neq i_2}{d(x_i, g(\vec{x}))}.
\end{align*}
By a previous result,
\begin{align*}
    A + d(x_{i_1}, c(\vec{x})) \leq \sqrt{2} B.
\end{align*}
Hence, it follows that
\begin{align*}
    [A + d(x_{i_1}, c(\vec{x}))] d(x_{i_2}', g(\vec{x})) \leq \sqrt{2} B d(x_{i_2}', g(\vec{x})).
\end{align*}
But since $b_g(\vec{x}) \geq a_g(\vec{x})$, it follows that $d(x_{i_2}', g(\vec{x})) \leq \sqrt{2} a_g(\vec{x})$. Hence,
\begin{align*}
    [A + d(x_{i_1}, c(\vec{x}))] d(x_{i_2}', g(\vec{x})) &\leq 2 B a_g(\vec{x})\\
    &= B (d(x_{i_1}', c(\vec{x})) - d(x_{i_2}', c(\vec{x}))).
\end{align*}
From this it follows that
\begin{align*}
    (A + d(x_{i_1}, c(\vec{x}))) (B + d(x_{i_2}', g(\vec{x}))) &= AB + B d(x_{i_1}, c(\vec{x})) + [A + d(x_{i_1}, c(\vec{x}))] d(x_{i_2}', g(\vec{x}))\\
    &\leq AB + B d(x_{i_1}', c(\vec{x}))\\
    &= (A + d(x_{i_1}', c(\vec{x}))) B
\end{align*}
and hence
\begin{align*}
    AR(\vec{x}) &= \frac{A + d(x_{i_1}, c(\vec{x}))}{B}\\
    &\leq \frac{A + d(x_{i_1}', c(\vec{x}))}{B + d(x_{i_2}', g(\vec{x}))}\\
    &= \frac{A + d(x_{i_1}', c(\vec{x}'))}{B + d(x_{i_2}', g(\vec{x}))}\\
    &\leq AR(\vec{x}').
\end{align*}
\end{proof}

\begin{lem}[Geometric to axis]
\label{Geometric to axis}
Suppose that $\vec{x}$ is a profile such that there are $a \geq 0$ and $b, c > 0$ and subsets $L, R, U \subseteq N$ with $L \cap R = L \cap U = R \cap U = \emptyset$, $L \cup R \cup U = N$, $|L| = 1$, $|U| = |R| = m$, and
\begin{itemize}
    \item $x_i = (-a, 0)$ for $i \in L$
    \item $x_i = (0, b)$ for $i \in U$
    \item $x_i = (c, 0)$ for $i \in R$
\end{itemize}
and so that $a_g(\vec{x}) > 0$.

% Let $\vec{x}'$ be the profile which is the same as $\vec{x}$ for $i \notin U$ and which has $x_i' = g(\vec{x})$ for $i \in U$. Then $AR(\vec{x}') \geq AR(\vec{x})$.

Then, there exists another profile $\vec{z}$ such that
$AR(\vec{z}) > AR(\vec{x})$.
\end{lem}

\begin{proof}

We'll consider two separate cases. \\

First assume that $a>0$. Let $\vec{x}(t)$ be the profile which is the same as $\vec{x}$ for $i \notin U$ and which has $\vec{x}_i(t) = g(\vec{x})+t\left(-a_g(\vec{x}),b-b_g(\vec{x})\right)$ for $i \in U$.
Then there exists $\epsilon > 0$ such that for $t \in [0, 1+\epsilon]$, 
\begin{align*}
    AR(\vec{x}(t)) &= \frac{(a + (1-t)a_g(\vec{x})) + m(c - (1-t)a_g(\vec{x})) + mb_g(\vec{x})+mt(b-b_g(\vec{x}))}{d((-a, 0), g(\vec{x})) + m d((c,0), g(\vec{x})) + mt d((0, b), g(\vec{x}))}\\
    &=\dfrac{t\left((m-1) a_g(\vec{x})+m(b-b_g(\vec{x}))\right)+a+a_{g}(\x)+m(c-a_{g}(\x)+b_g(\x))}{tmd((0, b), g(\vec{x}))+d((-a, 0), g(\vec{x})) + m d((c,0), g(\vec{x}))}
\end{align*}
Note that $\x(1)=\x$. Now since the denominator of $AR(\vec{x}(t))$ is strictly positive for $t \geq 0$ and since both the numerator and the denominator are linear in $t$, $AR(\vec{x}(t))$ is monotonic on $[0, 1 + \epsilon]$. There are three possibilities.\\

If $AR(\vec{x}(t))$ is strictly increasing, then $AR(\vec{x}(1+\epsilon)) > AR(\vec{x})$.

If $AR(\vec{x}(t))$ is strictly decreasing, then $AR(\vec{x}(0)) > AR(\vec{x})$.

If $AR(\vec{x}(t))$ is constant, consider the profile $\vec{z}'$ obtained by putting $t$ such that $-a=(1-t)a_g(\x)$. Then, under $\vec{z}'$, we have $1$ agent at $(-a,0)$, $m$ agents at $(-a,bt+(1-t)b_g(\x))$ and $m$ agents at $(c,0)$. Also, $AR(\x)=AR(\vec{z}')$. We can translate this profile by $a$ to the right and get a profile of points in the case where essentially we have $a=0$. We'll deal with this case now.\\

Now let's consider the case where $\x$ is such that $a=0$. To begin, note that since $g(\x)$ must be in the convex hull of $(0, 0)$, $(0, b)$ and $(c, 0)$, if $a_g(\x) \geq \frac{c}{2}$ and $b_g(\x) \geq \frac{b}{2}$, then $g(\x)=(\frac{c}{2}, \frac{b}{2})$. But then $\sum_{i=1}^n \dfrac{x_i-g(\x)}{\norm{x_i-g(\x)}}=\dfrac{(c,b)}{\sqrt{c^2+b^2}} \neq 0$, contradicting the characterization of the geometric median in Lemma~\ref{lem:gm}. Hence, it must be that either $a_g(\x)<\frac{c}{2}$ or $b_g(\x) < \frac{b}{2}$.

Suppose that $a_g(\x)<\frac{c}{2}$. Let $\vec{z}$ be the profile obtained from $\vec{x}$ by moving the point at $(0, 0)$ to $(a_g(\x) - \frac{c}{2}, 0)$ and moving one of the points at $(c, 0)$ to $(\frac{c}{2} + a_g(\x), 0)$. This transformation leaves coordinate-wise median unchanged, as well as leaving the sum of distances to the coordinate-wise median unchanged. However, the sum of distances to $g(\x)$ strictly decreases, since for the unaltered points the distance to $g(\x)$ remains the same, and the sum of the distances from the altered points to $g(\x)$ is
\begin{align*}
    d((0, 0), g(\x)) + d((c, 0), g(\x)) &= d((2a_g(\x), 0), g(\x)) + d((c, 0), g(\x))\\
    &> 2d((a_g(\x) + \frac{c}{2}), g(\x))\\
    &= d((a_g(\x) - \frac{c}{2}), g(\x)) + d((a_g(\x) + \frac{c}{2}), g(\x)),
\end{align*}
where the inequality follows from convexity of $d(\cdot, g(\x))$. Since $AR(\vec{z})$ is bounded below by the ratio of the sum of distances to the coordinate-wise median to the sum of distances to $g(\x)$, it follows that $AR(\vec{z}) > AR(\x)$.

Next, suppose that $b_g(\x) < \frac{b}{2}$. Let $\vec{z}$ be the profile obtained from $\vec{x}$ by moving the point at $(0, 0)$ to $(0, b_g(\x) - \frac{b}{2})$ and moving one of the points at $(0, b)$ to $(0, \frac{b}{2} + b_g(\x))$. By essentially the same argument just given, $AR(\vec{z}) > AR(\x)$.

% Note that a function $\frac{at+b}{ct+d}$ is decreasing iff $\frac{a}{c}<\frac{b}{d}$. So to show that $h(1) \leq h(0)$, we want to show that 

% $$\dfrac{(m-1) a_g(\vec{x})+m(b-b_g(\vec{x}))}{md((0, b), g(\vec{x}))} \leq \dfrac{a+a_{g}(\x)+m(c-a_{g}(\x)+b_g(\x))}{d((-a, 0), g(\vec{x})) + m d((c,0), g(\vec{x}))} $$

% Let $\alpha=a_g(\vec{x}), \beta=b_g(\vec{x}), d_1=d((-a,0),g(\x)), d_2=d((0,b),g(\x)), d_3=d((c,0),g(\x))$.

% Then, the above inequality can be written as
% $$\left(\alpha+m(\beta-\alpha)\right)\left(d_1+md_2+md_3\right)+amd_2-bmd_1+m^2cd_2-m^2bd_3 \geq 0$$

% Now, note that since the approximation ratio is always at least $1$, $h(0) = AR(\vec{x}') \geq 1$. Further,
% \begin{align*}
%     \lim_{t \to \infty}{h(t)} &= \frac{(m-1) a_g(\vec{x})+m(b-b_g(\vec{x}))}{m d((0, b), g(\vec{x}))}\\
%     &=\frac{((a_g(\vec{x})-0)+(b-b_g(\vec{x}))-\frac{1}{m}a_g(\vec{x})}{d((0, b), (a_g(\vec{x}),b_g(\vec{x})))}\\
%     &< \frac{a_g(\vec{x})}{d((0, b), g(\vec{x}))}\\
%     &< 1.
% \end{align*}
% Hence, there is some $t > 0$ such that $h(t) < 1 \leq h(0)$, and so since $h(t)$ is monotonic on $[0, \infty)$, it follows that $h(t)$ is decreasing on $[0, \infty)$. Thus, $AR(\vec{x}') = h(0) \geq h(1) = AR(\vec{x})$.
\end{proof}

Finally, the following lemma shows that we can use convexity to make the triangle formed by the three groups of points isosceles.

\begin{lem}[Isosceles]
\label{Isosceles}
Let $\vec{x}$ be a profile such for which are $m$ points at $(a,0)$, $1$ point at $(-b,0)$ and $m$ points at $(0,c)$, and for which $g(\vec{x})=(0,c)$ and $c(\vec{x})=(0,0)$. Let $\vec{x}'$ be the profile where there are $m$ points at $\left(\dfrac{ma+b}{m+1},0\right)$, $1$ point at $\left(-\dfrac{ma+b}{m+1},0\right)$, and $m$ points at $(0,c)$. Then, $AR(\vec{x}') \geq AR(\vec{x})$.
\end{lem}

\begin{proof}

Note that $c(\vec{x})=c(\vec{x}')=(0,0)$. Since $m*a+b=m*\frac{(ma+b)}{m+1}+\frac{ma+b}{m+1}$, we get that the numerator in $AR(\vec{x})$ and $AR(\vec{x}')$ remains the same. Thus, we only need to argue that the denominator goes down as we go from $AR(\vec{x})$ to $AR(\vec{x}')$.

Even though $g(\vec{x}')$ may not be equal to $g(\vec{x})$ we have that $sc(g(\vec{x}),\vec{x}') \leq  sc(g(\vec{x}),\vec{x})$ by the convexity of the distance function which would imply $sc(g(\vec{x}'),\vec{x}') \leq sc(g(\vec{x}),\vec{x})$ by definition of $g(\vec{x})$. Thus, we have that $AR(\vec{x}') \geq AR(\vec{x})$.

\end{proof}

Now, we use above lemmas to reduce the search space to I-CP.

\begin{lem}[ICP]
\label{icp2}
For every $\vec{x} \in CP$, there exists $\chi \in I-CP$ such that $AR(\chi) \geq AR(\vec{x})$.
\end{lem}

\begin{proof}
Without loss of generality, consider any profile $\vec{x} \in CP$ such that $b_g(\vec{x}) \geq a_{g}(\vec{x}) \geq 0$. 
Applying Lemma \ref{Reduce axes} to all points on the negative b axis gives a profile $\vec{x}'$  with a weakly higher approximation ratio. In $\vec{x}'$, we have all points on positive a, negative a, positive b and the geometric median. Using lemma  \ref{convexity}, we can combine the points on positive a, negative a, positive b to some $(a,0), (0,b), (-c, 0)$ while weakly increasing AR. Let this profile be $\vec{x}''$. Now, we use lemma \ref{Double Rotation} to move points on the geometric median to $+b$-axis. Using \ref{convexity} again,  we get a profile $\vec{x}'''$ with $m$ points on some $(a,0)$, 1 point on $(-c, 0)$ and $m$ points on $(0,b)$. 

So we know that there must be a worst-case profile that takes this form. From Lemma \ref{Geometric to axis}, we can say that if the geometric median of such a profile is not on the $b$-axis, it cannot be a worst-case profile. Thus, there must be a worst-case profile $\vec{z}$ with $m$ points on some $(a,0)$, 1 point on $(-c, 0)$ and $m$ points on $(0,b)$ and $a_g(\vec{z})=0$. Further, such a profile must have $b_g(\vec{z}) = b$, since otherwise, the profile with $m$ points on $(a, 0)$, 1 point on $(-c, 0)$, and $m$ points on $(0, b_g(\vec{z}))$ would have a strictly higher approximation ratio than $\vec{z}$. By Lemma~\ref{Isosceles}, since $\vec{z}$ is a worst-case profile, it must be that $c = a$.

Now, since $\vec{z}$ is a worst-case profile, the profile $\chi$ with $\chi_i = \frac{1}{b} \vec{z}_i$ is also a worst-case profile, and since $\chi \in I-CP$, the result follows.
\end{proof}

Using Lemma~\ref{icp2}, we can now restrict attention to profiles in $I-CP$. Define
\begin{align*}
    \vec{\eta}_t = (x_1^t, \dots, x_{2m+1}^t),
\end{align*}
where
\begin{align*}
    x_i^t = 
    \begin{cases} 
      (t, 0) & i = 1, \dots, m \\
      (-t, 0) & i = m+1 \\
      (0, 1) & i = m+2, \dots, 2m+1 
   \end{cases}
\end{align*}
Then,  $I-CP = \left \{ \vec{\eta}_t \, : \, t \geq \sqrt{\frac{2m + 1}{2m - 1}} \right \}$.
Defining $\a(t) = \frac{(m+1)t + m}{(m+1) \sqrt{t^2 + 1}}$, we get that for $t \geq \sqrt{\frac{2m + 1}{2m - 1}}$, $AR(\vec{\eta}_t) = \a(t)$, and that $\a(t)$ is maximized at $t^* = \frac{m+1}{m} > \sqrt{\frac{2m + 1}{2m - 1}}$, from which it follows that
\begin{align*}
    \text{Approximation ratio of CM} = \a \left( \frac{m+1}{m} \right ) = \sqrt{2}\dfrac{\sqrt{(2m+1)^2+1}}{(2m+1)+1} = \sqrt{2}\dfrac{\sqrt{n^2+1}}{n+1}.
\end{align*}

Thus, we get that the worst case approximation ratio is $\sqrt{2}\dfrac{\sqrt{n^2+1}}{n+1}$ as required.
\end{appendixproof}

For $n$ odd\footnote{
When $n=2m$ is even, the version of the coordinate-wise median mechanism given by $c(\vec{x}) = (\text{median}(-\infty,\vec{a}),\text{median}( -\infty,\vec{b}))$ has worst-case approximation ratio \textit{equal} to $\sqrt{2}$. This follows from the bound in Lemma ~\ref{bound} and the worst-case profile $\vec{x}$ where $x_1=x_2 \dots x_m=(1,0)$ and $x_{m+1}=x_{m+2} \dots x_{2m}=(0,1)$.}, the geometric median $g(\x)$ is unique and $OPT(\x) = g(\x)$. Hence, Theorem~\ref{minisum} amounts to finding how well the social cost of the coordinate-wise median mechanism approximates the social cost of the geometric median in the worst case.\\

The argument for obtaining the exact value of $AR(CM)$ is rather involved. We provide a full proof for the case that $n=3$ as we find the approach taken in its proof to be simple enough to be digestible yet sufficiently similar to the more nuanced approach required for arbitrary odd $n$ as to be illuminating. We then provide a sketch of the proof for all odd $n$, relegating the formal proof for this case to the appendix.\\

In both the $n=3$ case and the general case, the key to the proof is to reduce the search space for the worst-case profile from $(\mathbb{R}^2)^n$ to a much smaller space of profiles that have a simple structure. In many cases, this involves ``transforming" one profile into another profile that has a higher approximation ratio and a simpler structure. One important transformation that helps in significantly reducing the search space involves moving a point $x_i$ directly towards $g(\vec{x})$, getting as close as possible to $g(\vec{x})$ without changing $c(\vec{x})$. Because this transformation will be used repeatedly throughout this section, we provide here a proof that this transformation leads to a profile $(x_i', x_{-i})$ with a weakly higher approximation ratio.

\begin{lem}[Towards geometric median]
\label{togm}
Let $\vec{x}$ be a profile and $i \in N$, and let $\vec{x}'$ be any profile such that
\begin{enumerate}
    \item $x_i' \in [x_i, g(\vec{x})]$,
    \item for all $j \neq i$, $x_j' = x_j$, and
    \item $c(\vec{x}') = c(\vec{x})$.
\end{enumerate}
Then $AR(\vec{x}') \geq AR(\vec{x})$ where $
A R(\vec{x})=\frac{sc(c(\vec{x}), \vec{x})}{sc(g(\vec{x}), \vec{x})}
$
\end{lem}
\begin{proof}
By corollary~\ref{cor:gm}, $g(\vec{x}')=g(\vec{x})$ and by definition, $c(\vec{x}')=c(\vec{x})$. The change in optimal social cost is given by $\norm{x_i-x_i'}$ while the change in social cost with respect to coordinate-wise median is $\norm{c(\vec{x})-x_i'}-\norm{c(\vec{x})-x_i}$. By triangle inequality,  $\norm{x_i-x_i'} \geq \norm{c(\vec{x})-x_i'}-\norm{c(\vec{x})-x_i} $. Thus, the $sc(OPT(\cdot),\cdot)$ reduces by a greater amount than $sc(CM(\cdot), \cdot)$ as we move $x_i$ to $x_i'$. Since the ratio is always at least $1$, it follows that $AR(\vec{x}') \geq AR(\vec{x})$.

\end{proof}

\subsection{Proof for $n=3$ case}
\begin{corollary}
For $n=3$, the worst-case approximation ratio for the coordinate-wise median mechanism is given by: $$AR(CM) = \frac{\sqrt{5}}{2}.$$
\end{corollary}

\begin{remark}
\label{toricelli}
There is a more explicit characterisation of the geometric median when $n=3$. In this case, if any angle of the triangle formed by the three points is at least $120^o$, $g(\vec{x})$ lies on the vertex of that angle; otherwise, it is the unique point inside the triangle that subtends an angle of $120^o$ to all three pairs of vertices
\end{remark}

\begin{proof}[Proof of Theorem \ref{minisum} for $n=3$]
Define the set of \textit{Centered perpendicular (CP)} profiles as follows:
$$CP=\{\vec{x} \in (\mathbb{R}^{2})^3: c(\vec{x})=(0,0) \text{ and }  \forall i, \text{ either } a_i=0 \text{ or } b_i=0\}.$$
In words, a profile is in $CP$ if the  coordinate-wise median is at the origin and all points in $\vec{x}$ are on the axes.

Define the set of \textit{Isosceles-centered perpendicular (I-CP)} profiles as follows: $$I-CP=\{\vec{x} \in CP: \exists t \text{ such that } \vec{x}=((t,0), (-t, 0), (0,1)) \text{ and } g(\vec{x})=(0,1)\}$$ 
In words, a profile is in $I-CP$ if there are two points on the $a$-axis equidistant from the origin and the third point is at $(0,1)$, which is also the geometric median.

We first show that we can reduce the search space for the worst-case profile from  $(\mathbb{R}^2)^3$ to $CP$.

 \begin{lem}
 \label{lem:cp}

 For any profile $\vec{x} \in (\mathbb{R}^2)^3$, there is a profile $\vec{\chi} \in CP$ such that $AR(\vec{\chi}) \geq AR (\vec{x})$.
 \end{lem}
 
 \begin{proof}
 Let $\vec{x} \in (\mathbb{R}^2)^3$ be a profile. Let $\vec{x}'$ be the profile where $x_i' = x_i - c(\vec{x})$. Then $\vec{x}'$ has the same approximation ratio as $\vec{x}$ and $c(\vec{x}') = (0, 0)$. Denote $A = \{i: a_i=0\}$ and $B = \{i: b_i=0\}$. Note that since $c(\vec{x}') = (0, 0)$, it follows from the definition of $c(\vec{x}')$ that $A \neq \emptyset$ and $B \neq \emptyset$. For each $i$, define $x_i''$ as follows. Let $\G = \{(a, b) \in \R^2 \, : \, a = 0 \text{ or } b = 0\}$. If $i \in A \cup B$, let $x_i'' = x_i'$; otherwise, let $x_i''$ be the point in $[x_i', g(\vec{x}')] \cap \G$\footnote{The set $[x_i', g(\vec{x}')] \cap \G$ is non-empty because $g(\vec{x}')$ cannot be in the same quadrant as $x_i'$. Any point in the same quadrant as $x_i'$ subtends an angle of less than $90^o$ with the other two points and hence it cannot be the geometric median.} that is closest to $x_i'$.  Then $x_i'' \in \G$ for all $i$ and $c(\vec{x}'') = (0, 0)$, so $\vec{x}'' \in CP$. Further, it follows from Lemma~\ref{togm} that $AR(\vec{x}'') \geq AR(\vec{x}') = AR(\vec{x})$; hence, taking $\vec{\chi} = \vec{x''}$ completes the proof.

 \end{proof}

 \begin{tikzpicture}[scale=0.7][H]
\label{fig:move}

\filldraw (1,3) circle[radius=2pt];
\filldraw (4,-1) circle[radius=2pt];
\filldraw (-2,-3) circle[radius=2pt];

\filldraw[blue] (1,-1) circle[radius=2pt];
\filldraw[red] (1.523, -0.08) circle[radius=2pt];
\filldraw (0.4,-1) circle[radius=2pt];

\draw [green] (1,3) -- (1,-4);
\draw [green] (4,-1) -- (-2,-1);
\draw (-2,-3) -- (1.523, -0.08);
\draw [dashed] (-2,-3) -- (1,-1);

\node[left=1pt of {(1,3)}, outer sep=2pt] {$x_2$};
\node[right=1pt of {(4,-1)}, outer sep=2pt] {$x_1$};
\node[left=1pt of {(-2,-3)}, outer sep=2pt] {$x_3$};

\node[below right=1pt of {(1,-1)}, outer sep=2pt] {$c(\vec{x})$};
\node[above right=1pt of {(1.523, -0.08)}, outer sep=2pt] {$g(\vec{x})$};
\node[above=1pt of {(0.5,-1)}, outer sep=2pt] {$x_3'$};

\node[below=0.5pt of {(0,-4)}, outer sep=2pt] {Figure \ref{fig:move}: Towards geometric median};
\end{tikzpicture}
 
 Now we show that we can further reduce the search space from $CP$ to $I-CP$.

 \begin{lem}
  \label{lem:icp}

 For any profile $\vec{x} \in CP$, there exists a profile $\vec{\chi} \in I-CP$ such that $AR(\vec{\chi}) \geq AR (\vec{x})$.
 \end{lem}
\begin{proof}

Let $\vec{x}$ be a profile in $CP$. 

Without loss of generality, we may assume that all $x_i$ are weakly above the $a$-axis and there are at least two $x_i$ on the $a$-axis, since reflecting a profile in $CP$ across the $a$-axis, the $b$-axis, or the line $a = b$ gives a profile in $CP$ with the same approximation ratio. Hence, we can label the points such that $x_1 = (-a, 0)$, $x_2 = (b, 0)$, and $x_3 = (0, c)$, for some $a, b, c \geq 0$. 

If $c = 0$, then $AR(\vec{x}) = 1$, and so every profile has approximation ratio weakly greater than $\vec{x}$. Hence, we may further assume that $c > 0$. 

Since $x_1$ and $x_2$ are on the $a$-axis, it follows from the characterization of the geometric median for three points given in remark~\ref{toricelli} that $-a \leq a_g(\vec{x}) \leq b$ and $0 < b_{g}(\vec{x}) \leq c$. Hence, moving $x_3$ to $g(\vec{x})$ then (if necessary) translating all points by the same vector so that the coordinate-wise median is at the origin yields a profile in $CP$ which has higher approximation ratio. Hence, we may further assume that $g(\vec{x}) = x_3$.

Let $\vec{x}'$ be the profile where $x_1' = (-(a + b)/2, 0)$, $x_2' = ((a + b)/2, 0)$, and $x_3' = (0, c)$. 
By definition, $sc(g(\vec{x}'), \vec{x}') \leq sc(g(\vec{x}), \vec{x}')$ and by an argument that exploits the convexity of the distance function, $sc(g(\vec{x}), \vec{x}') \leq sc(g(\vec{x}), \vec{x})$. Combining these inequalities gives $sc(g(\vec{x}'), \vec{x}') \leq sc(g(\vec{x}), \vec{x})$, and a simple calculation shows that $sc(c(\vec{x}'), \vec{x}')=sc(c(\vec{x}) ,\vec{x})$. Thus, $AR(\vec{x}') \geq AR(\vec{x})$.

Note that under $\vec{x}'$, $g(\vec{x}')=(0,k)$ for some $k \leq c$. Define $\vec{x}''$ to be the profile with  $x_1''=x_1'$, $x_2''=x_2'$, and $x_3''=g(\vec{x}')$. Then, by Lemma~\ref{togm}, $AR(\vec{x}'') \geq AR(\vec{x}')$.

Finally, define $\vec{x}'''$ such that $x_i''' = \frac{1}{c} x_i''$ for each $i$. Then since $AR(\cdot)$ is homogeneous of degree $0$, $AR(\vec{x}''') = AR(\vec{x}'')$, and so $AR(\vec{x}''') \geq AR(\vec{x})$.  Further, $c(\vec{x}''') = (0,0)$, $x_1''' = (-t, 0)$, $x_2''' = (t, 0)$, and $x_3''' = (0, 1)$ for some $t \geq 0$; in fact, it follows from the characterisation of the geometric median that $t \geq \sqrt{3}$. Hence, $\vec{x}''' \in I-CP$, and so taking $\vec{\chi} = \vec{x}'''$ completes the proof.

\end{proof}

Denote by $\vec{\eta}_t = ((t,0), (-t, 0), (0,1)).$ It follows from the arguments in the proof of Lemma~\ref{lem:icp} that $I-CP=\{\vec{\eta}_t \, : \, t \geq \sqrt{3}\}$. Let $\a(t) = \frac{2t + 1}{2 \sqrt{t^2 + 1}}$. A simple calculation shows that for $t \geq \sqrt{3}$, $AR(\vec{\eta}_t) = \a(t)$. In particular, it follows that the approximation ratio of coordinate-wise median mechanism is equal to $\sup_{t \geq \sqrt{3}}{\a(t)}$. Since $\a(t)$ achieves its global maximum at $t^* = 2 > \sqrt{3}$, the ratio is $ AR(\vec{\eta}_2) = \a(2)$. Since $\a(2) = \sqrt{2}\dfrac{\sqrt{3^2+1}}{3+1}$, the result follows. 

\end{proof}

\subsection{Proof sketch for general $n$}

\begin{proofsketch}

We now consider the case of $n=2m+1$ agents. We begin by defining classes of profiles analogous to those used in the proof for $n=3$. 

We define the class of Centered Perpendicular (CP) profiles as all profiles $\vec{x} \in (\mathbb{R}^2)^n$ such that 

\begin{itemize}
    \item $c(\vec{x})=(0,0)$
    \item for all $i$, either $a_i=0$ or $b_i=0$ or $x_i=g(\vec{x})$
    \item if $x_i' \in (x_i, g(\vec{x}))$, then $c(x_i', x_{-i}) \neq (0,0)$
\end{itemize}

Since the last condition is slightly more subtle than the others and will be important in the sequel, we describe it now in words. This condition says that \textit{any} (nonzero) movement of \textit{any} $x_i$ towards the geometric median would result in a change in the coordinate-wise median.

We define the class of Isosceles-Centered Perpendicular (I-CP) profiles as all $\vec{x} \in CP$ for which there exists $t \geq 0$ such that
\begin{itemize}
    \item $x_1= \dots =x_m=(t,0)$
    \item $x_{m+1}=(-t, 0)$
    \item $x_{m+2}= \dots= x_{2m+1}=(0,1)$
    \item $g(\vec{x})=(0,1)$.
\end{itemize}

The proof proceeds much as in the proof for $n=3$. We first show that for every profile, there is some profile in $CP$ with weakly higher approximation ratio. The approach used in the $n=3$ case extends naturally here: first, translate the profile $\vec{x} \in (\R^2)^n$ so that coordinate-wise median moves to the origin; then, starting from $i = 1$ and going to $i = n$, move $x_i$ directly towards the geometric median until either it reaches the geometric median or moving it further would move the coordinate-wise median. The resulting profile is in $CP$ and has an approximation ratio that is weakly greater than $\vec{x}$'s.

\begin{center}

\qquad
\begin{tikzpicture}[scale=0.7]

\label{fig:npoints}

\draw[green] (-3,0) -- (3,0);
\draw[green] (0,-3) -- (0,3);

\filldraw (1,0) circle[radius=2pt];
\filldraw (2,0) circle[radius=2pt];
\filldraw (-0.8,0) circle[radius=2pt];
\filldraw (0,0.8) circle[radius=2pt];
\filldraw (0,1.4) circle[radius=2pt];
\filldraw (0,-0.5) circle[radius=2pt];
\filldraw (0,-2) circle[radius=2pt];
\filldraw (-2,0) circle[radius=2pt];

\filldraw[black] (1,1) circle[radius=4pt];

\filldraw[red] (1,1) circle[radius=2pt];
\filldraw[blue] (0,0) circle[radius=2pt];

\node[below right=1pt of {(0,0)}, outer sep=2pt] {$c(\vec{x})$};
\node[right=1pt of {(1, 1)}, outer sep=2pt] {$g(\vec{x})$};

\node[below right=1pt of {(-3, -3)}, outer sep=2pt] {Figure \ref{fig:npoints}: A CP profile};

\end{tikzpicture} 
\end{center}

Next, we show that for any profile in $CP$, there is some profile in $I-CP$ with weakly higher approximation ratio. The approach used in the $n=3$ case for this step \textit{does not} extend in a straightforward manner to the general case---the main obstruction arises from the fact that for a profile $\vec{x}$ in $CP$, there may be $i \in N$ such that $x_i = g(\vec{x})$, which may not be on either axis. The next subsection is devoted to giving an overview of the procedure used to transform a profile in $CP$ to one in $I-CP$ with weakly higher approximation ratio.

Finally, the approach used to calculate the worst-case approximation ratio for profiles in $I-CP$ has much the same structure as in the $n=3$ case. We define $\vec{\eta}_t = (x_1^t, \dots, x_{2m+1}^t),$
where
\begin{align*}
    x_i^t = 
    \begin{cases} 
      (t, 0), & i = 1, \dots, m \\
      (-t, 0), & i = m+1 \\
      (0, 1), & i = m+2, \dots, 2m+1 
   \end{cases}
\end{align*}
and we show that $I-CP = \left \{ \vec{\eta}_t \, : \, t \geq \sqrt{\frac{2m + 1}{2m - 1}} \right \}$.
Defining $\a(t) = \frac{(m+1)t + m}{(m+1) \sqrt{t^2 + 1}}$, we show that for $t \geq \sqrt{\frac{2m + 1}{2m - 1}}$, $AR(\vec{\eta}_t) = \a(t)$, and that $\a(t)$ has a global maximum at $t^* = \frac{m+1}{m} > \sqrt{\frac{2m + 1}{2m - 1}}$, from which it follows that
\begin{align*}
    AR(CM) = \a \left( \frac{m+1}{m} \right ) = \sqrt{2}\dfrac{\sqrt{(2m+1)^2+1}}{(2m+1)+1} = \sqrt{2}\dfrac{\sqrt{n^2+1}}{n+1}.
\end{align*}

\begin{center}
\begin{tikzpicture}[scale=0.8]
\label{fig:worst}

\draw[green] (-3,0) -- (3,0);
\draw[green] (0,-3) -- (0,3);

\filldraw[black] (3,0) circle[radius=4pt];

\filldraw[black] (0,2) circle[radius=4pt];
\filldraw[red] (0,2) circle[radius=2pt];

\filldraw (-3,0) circle[radius=2pt];

\node[below=1pt of {(-3,0)}, outer sep=2pt] {$(-\frac{m+1}{m},0)$};
\node[below=1pt of {(3,0)}, outer sep=2pt] {$(\frac{m+1}{m},0)$};

\node[left=1pt of {(0,2)}, outer sep=2pt] {$(0,1)$};

\node[above=1pt of {(3,0)}, outer sep=2pt] {$m$ agents};

\node[above=1pt of {(-3,0)}, outer sep=2pt] {$1$ agent}; 

\node[above right=1pt of {(0, 2)}, outer sep=2pt] {$m$ agents};

\filldraw[blue] (0,0) circle[radius=2pt];

\node[below right=1pt of {(0,0)}, outer sep=2pt] {$c(\vec{x})$};
\node[below right=1pt of {(0, 2)}, outer sep=2pt] {$g(\vec{x})$};

\node[below right=1pt of {(-3, -3)}, outer sep=2pt] {Figure \ref{fig:worst}: Worst case profile};
\end{tikzpicture} 
\end{center}

\end{proofsketch}

\subsection{Reduction from CP to ICP}

In this subsection,  we discuss informally some transformations that allow us to deal with the profiles in $CP$. Without loss of generality (using reflections if necessary as in the $n=3$ case), we may restrict consideration to profiles $\vec{x} \in CP$ with $g(\vec{x})=(a_g,b_g)$ such that $a_g \geq 0$, $b_g \geq 0$, and $b_g \geq a_g$. 

\begin{enumerate}
    \item \textbf{Reducing axes}: In this step, we move all points on $-b$-axis to $-a$-axis while keeping them equidistant from $c(\vec{x})=(0,0)$. This works because the $sc(c(\cdot),\cdot)$ remains the same while $sc(g(\cdot),\cdot)$ reduces, as the points move closer to the old geometric median. Thus, we get a profile in which all points are either on one of the $+a$-, $+b$-, or $-a$-axes or at $g(\vec{x})$.
    
    \item \textbf{Convexity}: Consider a profile obtained after applying step 1. Transform the profile so that all points on the $+a$-, $+b$-, and $-a$-axes are at their mean coordinates on the $+a$-, $+b$-, and $-a$-axes respectively. Again, $sc(c(\cdot),\cdot)$ remains the same while $sc(g(\cdot),\cdot)$ falls because of convexity of the distance function. Thus, we get a profile with weakly higher approximation ratio which has $k$ points at $(-b, 0)$, $m+1-k$ points at $(0,c)$, $m+1-k$ points at $(a,0)$ and $k-1$ points at $g(x)$. Note that we are able to pin down the exact cardinalities of these sets because of the third condition in the definition of $CP$, which requires that if any of the points were to move towards $g(\vec{x})$, then $c(\vec{x})$ would change.
    
    \item \textbf{Double Rotation}: Consider a profile obtained after applying step 2. Transform the profile by moving the $k-1$ points at $g(\vec{x})$ to $(0, \alpha)$, where $\alpha=d(c(\vec{x}),g(\vec{x}))$, and moving $k-1$ of the $k$ points at $(-b,0)$ to $(\beta, 0)$, where $\beta$ is the unique positive number such that $d(g(\vec{x}), (\beta, 0))=d(g(\vec{x}), (-b,0))$. In this case, one can show that the increase in $sc(c(\cdot),\cdot)$ is at least $\sqrt{2}$ times the increase in $sc(g(\cdot),\cdot)$ and therefore,  by Lemma~\ref{bound}, it follows that the approximation ratio weakly increases. Applying convexity again, we get a profile such that there is one point at $(-b, 0)$, $m$ points at $(0,c)$ and $m$ points at $(a,0)$. Note that $g(\vec{x})$ may still not be on the axes.
    
    \item \textbf{Geometric to axis}: Consider a profile obtained after applying step 3. 
    In the case that $g(\vec{x})$ is not on the axes, we show that moving the $m$ points at $(0,c)$ directly towards or away from  $g(\vec{x})$ strictly increases the ratio. It follows then that there must be a worst-case profile where one point is at $(-b,0)$, $m$ points are at $(0,c)$, $m$ points are at $(a,0)$ and $g(\vec{x})=(0,c)$.
\end{enumerate}

From here, we apply a transformation similar to step 2 to get a profile in $I-CP$. Note that we have suppressed some details (especially when the same transformation must be used repeatedly) in order to make the exposition as clear as possible---see the appendix for a rigorous proof.

%% file: files/p-norm.tex
\section{p-norm objective}

In this section, we consider the problem of quantifying the approximation ratio for the coordinate-wise median mechanism under the $p$-norm objective $sc(y,\vec{x})=(\sum_{i=1}^n \norm{y-x_i}^p)^\frac{1}{p}$
for $p \geq 2$. While we do not exactly quantify the AR for arbitrary $n$ in this case, we are able to obtain bounds on the asymptotic AR of the coordinate-wise median mechanism.

\begin{thmrep}
\label{2d}
For $X=\mathbb{R}^2$ and the $p$-norm objective with $p \geq 2$, 
$$2^{1-\frac{1}{p}} \leq \sup_{n \in \mathbb{N}} AR(CM) \leq 2^{\frac{3}{2}-\frac{2}{p}}$$
\end{thmrep}

\begin{appendixproof}

We'll prove that the lower bound actually holds for any deterministic, strategyproof mechanism (defined for all $n \in \mathbb{N}$) and hence, it holds for the coordinate-wise median mechanism. So suppose $f$ is any deterministic, strategyproof mechanism. With $n=2m+1$ agents\footnote{The same argument applies if we just take $n=2$ but we wanted to illustrate that the result holds even under the restriction to odd number of agents}, for any profile $\x$ such that $m$ agents have ideal point $\alpha$, $m+1$ agents have ideal point $\beta \neq \alpha$, and $f(\x) = \beta$,
\begin{align*}
    AR_f(\x') &=\dfrac{sc(f(\x),\x)}{sc(OPT(\x),\x)}\\
    &\geq \dfrac{\left(m*\norm{\alpha - \beta}^p\right)^\frac{1}{p}}{\left((2m+1) (\norm{\alpha - \beta}/2)^p\right)^\frac{1}{p}}\\
    &= 2^{1-\frac{1}{p}} \cdot \left(\dfrac{m}{m + 1/2}\right)^\frac{1}{p} \cdot
\end{align*}
To see that such a profile always exists, consider the profile $\x$ where agents $1$ through $m$ have ideal point $(-1, 0)$ and agents $m+1$ through $2m+1$ have ideal point $(1, 0)$. If $f(\x) = (1, 0)$, then $\x$ is such a profile; if not, let $\x'$ be the profile where agents $1$ through $m+1$ have ideal point $f(\x)$ and agents $m+2$ through $2m+1$ have ideal point at $(1, 0)$. Since $\x'$, every agent's ideal point is either the same as under $\x$ or equal to $f(\x)$, it follows from strategyproofness that $f(\x') = f(\x)$, and so $\x'$ is such a profile.

Thus, for any $n = 2m+1$,
\begin{align*}
    AR(f) \geq 2^{1-\frac{1}{p}} \cdot \left(\dfrac{n-1}{n}\right)^\frac{1}{p},
\end{align*}
so
\begin{align*}
    \sup_n{AR(f)} \geq 2^{1-\frac{1}{p}}.
\end{align*}

Now we show that the asymptotic AR of coordinate-wise median mechanism is bounded above by $2^{\frac{3}{2}-\frac{2}{p}}$. Consider any profile $\vec{x}=(a_i, b_i) \in (\mathbb{R}^2)^n$. Let $g(\vec{x})=(a_g(\vec{x}), b_g(\vec{x}))$ and $c(\vec{x})=(a_c(\vec{x}),b_c(\vec{x}))$. Then, we have that 

\begin{align*}
    sc(g(\vec{x}), \vec{x})^p&=\sum_{i=1}^n \norm{g(\vec{x})-x_i}^p\\
    &\geq \left(\sum_{i=1}^n  \norm{a_g(\vec{x})-a_i}^p+\sum_{i=1}^n  \norm{b_g(\vec{x})-b_i}^p\right)\\
    &\geq \left(\sum_{i=1}^n  \norm{OPT(a)-a_i}^p+\sum_{i=1}^n  \norm{OPT(b)-b_i}^p\right)\\
    &\geq \dfrac{1}{2^{p-1}}\left(\sum_{i=1}^n  \norm{c_a-a_i}^p+\sum_{i=1}^n  \norm{c_b-b_i}^p\right)\\
    &\geq \dfrac{2^{1-\frac{p}{2}}}{2^{p-1}}\sum_{i=1}^n \norm{c(\vec{x})-x_i}^p\\
    &=2^{2-\frac{3p}{2}}sc(c(\vec{x}),\vec{x})^p
\end{align*}

Thus, we get $AR(CM) \leq 2^{\frac{3}{2}-\frac{2}{p}}$ for $p \geq 2$ as required.
\end{appendixproof}

The lower bound follows directly from Lemma \ref{1d}, since restriction of the coordinate-wise median mechanism to profiles on the $a$-axis corresponds to the median mechanism in one dimension. \footnote{The lower bound actually holds more generally in that if $f$ is a deterministic, strategyproof mechanism defined for all $n$, then $\sup_{n \in N} AR(f) \geq 2^{1-\frac{1}{p}}$. If $f$ is anonymous as well, the bound is a corollary of Theorem~\ref{2d} due to the optimality of Coordinate-wise median (Theorem~\ref{thm:cm-best}) for any $n$. For any $f$, the argument in \citet{feigenbaum_approximately_2017} to show Lemma~\ref{1d} extends to this setting as well and is in the appendix proof.
} 

For the upper bound, we again use Lemma \ref{1d} and note that, if $a_c$ is the median of $(a_1, a_2,\dots, a_n)$ and $OPT(a)$ is the optimal location, then 
\begin{align*}
    \sum_{i=1}^n \norm{a_c-a_i}^p \leq 2^{p-1} \sum_{i=1}^n \norm{OPT(a)-a_i}^p.
\end{align*}
The upper bound is then obtained by using the following inequalities, together with Lemma~\ref{1d}:
\begin{align*}
(\alpha^2+\beta^2)^{\frac{p}{2}} \geq (\alpha^p+\beta^p) \;\;\;\;\;\;\;\;\;\;\;\; \alpha^p+\beta^p \geq 2^{1-\frac{p}{2}}(\alpha^2+\beta^2)^{\frac{p}{2}}.
\end{align*}

The full proof is relegated to the appendix.\\

For $p=2$, the upper and lower bound in Theorem \ref{2d} coincide and we get the following:
\begin{corollary}
For $X=\R^2$ and $sc(y,\vec{x})=\left[\sum \norm{y-x_i}^2\right]^\frac{1}{2}$ ($p=2$),  $$\sup_{n \in \mathbb{N}} AR(CM) =\sqrt{2}$$
\end{corollary}

For $p=\infty$, any deterministic strategyproof mechanism has $AR  \geq 2$. Also, any Pareto optimal mechanism has $AR \leq 2$. Together, we get

\begin{corollary}
For $X=\R^2$ and $sc(y,\vec{x})=\max_i \norm{y-x_i}$ ($p=\infty$),  $$AR(CM)=2$$.
\end{corollary}

The last corollary suggests that the upper bound in Theorem \ref{2d} is not tight. In fact, the AR of CM is actually equal to its lower bound in both cases $p=2$ and $p=\infty$. This leads us to conjecture that:
\begin{conjecture}
\label{conj:CM-very-best}
For $X=\R^2$, and the $p$- norm objective $sc(y, \vec{x})=\left[\sum \norm{y-x_i}^p \right]^\frac{1}{p}$ where $p\geq 2$,  $$\sup_{n \in \mathbb{N}} AR(CM)=2^{1-\frac{1}{p}}$$.
\end{conjecture}

%% file: files/conclusion.tex
\section{Conclusion}

In this work, we demonstrate that the coordinate-wise median mechanism is the optimal deterministic, anonymous, and strategyproof mechanism for a large, natural class of social cost functions. We show that the utilitarian cost of the coordinate-wise median is always within $\sqrt{2}\frac{\sqrt{n^2+1}}{n+1}$ of the utilitarian  cost obtained under the optimal mechanism. For the $p$-norm objectives, we find that the worst-case approximation ratio for the coordinate-wise median mechanism is bounded above by $2^{\frac{3}{2}-\frac{2}{p}} $ for $p \geq 2$. For the case of $p=2$ and $p=\infty$, the coordinate-wise median mechanism has AR equal to $\sqrt{2}$ and $2$, respectively. This leads us to conjecture that the AR of coordinate-wise median mechanism is actually equal to $2^{1-\frac{1}{p}}$ for any $p\geq 2$. \\

% Work on higher-dimensional facility location problems remains hitherto relatively limited, but we are optimistic that the results and methods used in this paper will encourage further research in this fundamental domain. We wish to stress, in particular, that the simple structure of the worst-case profile can provide a ready tool for researchers looking to advance progress on the conjectural optimality of the coordinate-wise median mechanism.

We hope that the results and methods in this paper will encourage further research in this fundamental domain. The question of how well a randomized mechanism might approximate the social cost of the geometric median remains open. A potentially good candidate is the mechanism that chooses a coordinate-wise median after a uniform rotation of the orthogonal axes. While its analysis seems hard in general, finding its AR on the worst-case profile in Theorem~\ref{minisum} might give a useful lower bound. Another question is to close the gap between the upper bound on AR of the coordinate-wise median mechanism and the lower bound on AR of any deterministic strategyproof mechanism for the $p$ norm objective. The analysis for more general single-peaked preferences in multi-dimensional domains also remains open.